\documentstyle[11pt,newpasp,twoside,epsf]{article}

\reversemarginpar

\begin{document}
\title
{
 Synthetic Spectra from 3D Models of Supernovae
}
\author
{
 R. C. Thomas, E. Baron, David Branch
}
\affil
{
 University of Oklahoma
 Nielsen Hall RM 131
 Norman, Oklahoma 73019
}
\begin{abstract}
We describe \texttt{Brute}, a code for generating synthetic spectra
from 3D supernova models.  It is based on a Monte Carlo implementation
of the Sobolev approximation.  The current version is highly
parameterized, but in the future it will be extended for full NLTE
spectrum modelling of supernovae.
\end{abstract}

\section{Introduction}

The amount of evidence that the envelopes of supernovae (SNe) can 
deviate from spherical symmetry is increasing.  Net intrinsic polarization
measurements from some SNe are consistent with ellipsoidal envelopes
(Howell et al.\ 2001; Leonard et al.\ 2001; Leonard et al.\ 2000; Wang
et al.\ 2001; Wang et al.\ 1997; see Wheeler 2000 for a list of
measurements).  In some flux spectra, a clumpy or macroscopically
mixed ejecta distribution (or at least a nonspherical excitation 
structure) has been invoked to explain phenomena such as the ``Bochum
event'' in SN 1987A (Phillips \& Heathcote 1989; Utrobin, Chugai \&
Andronova 1995).  Most conspicuously, the interesting morphologies of
SN remnants may indicate macroscopic mixing in the initial SN
envelopes (Decourchelle et al.\ 2001; Fesen \& Gunderson 1996; Hwang,
Holt \& Petre 2000; Tsunemi, Miyama \& Aschenbach 1995).  These
findings have inspired new 3D explosion models (Khokhlov 2000;
Reinecke, Hillebrandt \& Niemeyer 2002; Kifonidis et al.\ 2000).
Testing new models as they arise will require detailed NLTE radiative
transfer calculations in 3D.

Nonetheless, considering and deducing general constraints on SN
geometry without such detailed calculations can be fruitful (Thomas et
al.\ 2002).  We have developed a parameterized synthetic spectrum code
for 3D models of SNe called \texttt{Brute} which we describe here.  In
\S 2 we describe the assumptions and implementation of \texttt{Brute},
and in \S 3 we 
outline future work to bring the code toward a detailed analysis
code.

\section{Model}

\texttt{Brute} was designed under some of the same basic assumptions
as the highly successful direct analysis code \texttt{Synow} (Fisher
2000; Deng et al.\ 2000; Millard et al.\ 1999).  Both codes use a
simplified model of a SN atmosphere which has been called the
``elementary supernova'' (ES; Jeffery \& Branch 1990).  The major
difference is that \texttt{Synow} is a strictly 1D code, but
\texttt{Brute} is for arbitrary geometry.

\subsection{The Elementary Supernova Model}
The ES model is a simplified picture of a SN at phases between a few
days to a few months after explosion.  It consists of a central,
optically thick continuum-emitting core surrounded by an extended
envelope within which line formation occurs.  The SN expands
homologously so that $v \propto r$ for all matter elements.

The ES core (or photosphere) is approximated by a surface which emits
blackbody radiation at some specified temperature.  The photosphere is
sharp, meaning that all radiation regardless of wavelength originates
at the same physical surface.

Radiation transfer in the envelope is accomplished by means of the
Sobolev approximation (Sobolev 1947; Rybicki \& Hummer 1978),
justified in detail for the SN case by Jeffery \& Branch (1990).  The
assumption of homologous expansion provides a great simplification in
the calculations, because then (1) Sobolev optical depth is angle
independent, (2) common point velocity surfaces for intensity
integrals are spheres and (3) propagating photons only redshift with
respect to the matter and never blueshift.  For each ion used in the
envelope, the optical depth of a ``reference line'' is parameterized
spatially using a contrast function.  All other line optical depths of
the same ion are scaled assuming LTE.  The source function is taken to
be that of pure resonance scattering ($S = J$).

It should be noted that the parameterized prescription here can be
replaced by one with more detail.  The resonance scattering assumption
can be generalized as well.  The effects of special relativity in the
transfer can be included (to minor effect; Jeffery 1993).  Electron
scattering also may be included.  To speed up transfer, lines close 
together in wavelength can be grouped into bands of small width
to prevent needless extra calculations.  For the purpose of direct
analysis of SN spectra however, the above parameterized 
prescription is sufficient.

Once the source function has been determined for each line at all
points in the envelope, a formal solution along impact parameter beams
toward an observer is calculated.  Again, the Sobolev approximation
simplifies this process, since emission in a given wavelength from the
SN arises only from common direction velocity planes perpendicular
to the line of sight.

\subsection{3D Implementation}

The key difference between the 1D and 3D implementation is the method
by which the mean intensity is calculated.  In 1D, we can take
advantage of an azimuthal symmetry in the intensity arriving at a
point to simplify the mean intensity integral.  But in 3D, such
symmetry is missing and the integral becomes more problematic.  To
avoid this problem, we adopt a Monte Carlo approach, based largely on
1D codes (Abbott \& Lucy 1985; Mazzali \& Lucy 1993; Mazzali 2000).

Packets of equal energy with wavelengths drawn from the Planck 
distribution are emitted from the surface of the core such that the
core is a Lambert radiator.  The packet random walks from line
scattering to line scattering as it redshifts into resonance with
lines of increasing wavelength.  The probability that a packet
undergoes a scattering (coherent in the comoving frame) is determined
by the optical depth at the resonance targets.  

As each packet enters a new cell of the envelope or immediately after a
scattering, an ``event'' optical depth is chosen to ``schedule'' the
next possible scattering : $\tau_{ev} = - \ln( R )$ where $R$ is a
random number between 0 and 1.  As the packet propagates through the
cell, it accumulates optical depth from the lines it encounters until
the total exceeds $\tau_{ev}$.  The packet then scatters coherently.

From the perspective of a single cell in the envelope during a 
simulation, packets of equal energy intersect the cell from many
directions.  The locus of points defined by the packets' previous
resonance targets form the common point velocity surfaces.  The Monte
Carlo technique assembles these surfaces implicitly.  The radiation
field is built up by counting packets that come into resonance with
lines everywhere in the envelope.  At the end of the simulation, the
total luminosity is used to calibrate the weight of a single packet,
converting the packet-in-resonance tallies into estimates of $J$ in
each line.

\begin{figure}
\plotfiddle{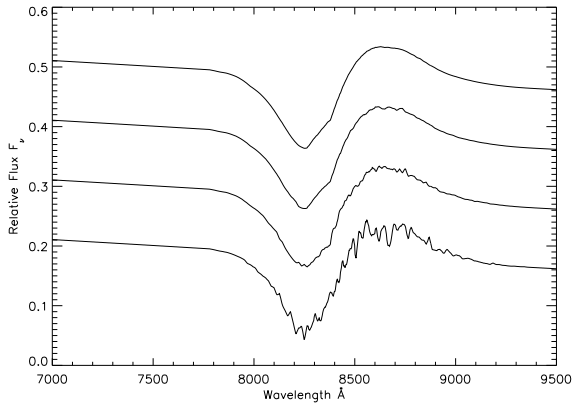}{2.0in}{0.0}{45}{45}{-124}{0}
\caption{The effect of increasing packet number on noise.  The Ca infrared 
triplet is plotted for (from bottom to top) $10^4$, $10^5$, $10^6$
and $10^7$ packets used.}
\end{figure}

Of course in 3D, the appearance of a spectrum may depend on the line
of sight to the envelope.  Rather than wait for enough packets to
emerge along all lines of sight (a serious problem), we employ the
same formal integral technique used in the 1D calculation.  Once the
source function is determined everywhere by Monte Carlo, common
direction plane positions are computed and the flux is integrated
across impact parameter beams.  This technique greatly reduces the 
number of packets to follow in a simulation.

\section{Future work}

In order to resolve variations in optical depth that are small in
size, the resolution of the envelope must be increased.  The 
simplest solution is to use a big computer with a lot of memory,
or split the envelope up amongst multiple processors.  The fact
that we need only sweep the line list from blue to red may help,
and we can load segments of the line list as the calculation
progresses.

Including electron scattering in the calculations, and replacing
the assumption of pure resonance scattering with branching will
help to make the code more consistent overall.  The eventual
goal is a NLTE code for use with new 3D explosion models.  Until
the reliability and quantity of those models increases, a great
deal about deviations from spherical symmetry in SNe can be 
learned from a parameterized code.

\acknowledgements

This work was supported in part by NASA grants NAG5-3505 and
NAG5-12127, and an IBM SUR grant to the University of Oklahoma.

\end{document}